\documentclass[pra,twocolumn,aps,showpacs,superscriptaddress]{revtex4}
\usepackage{graphicx}
\usepackage{multirow}
\usepackage{amsmath}
\usepackage{amssymb}
\usepackage{amsthm}
\usepackage{eucal}
\usepackage{bm}
\usepackage{upgreek}
\usepackage{framed}
\usepackage{mathrsfs}
\usepackage{latexsym}
\usepackage{float}
\usepackage{subfigure}
\usepackage{color}

\newcommand{\ket} [1] {\vert #1 \rangle}
\newcommand{\bra} [1] {\langle #1 \vert}

\newcommand{\hr}[1]{\hat{\rho}_{#1}}
\newcommand{\bs}[1]{\boldsymbol{#1}}
\newcommand{\id}[1]{1\!\!1_{#1}}

\begin{document}
\title{Process output nonclassicality and nonclassicality depth of quantum-optical channels}
\author{Krishna Kumar Sabapathy}
\email{krishnakumar.sabapathy@gmail.com}
\affiliation{F\'{i}sica Te\`{o}rica: Informaci\'{o} i Fen\`{o}mens Qu\`{a}ntics, Departament de F\'{i}sica, Universitat Aut\`{o}noma de Barcelona, 08193 Bellaterra (Barcelona), Spain.}

\begin{abstract}
 We introduce a quantum-optical notion of nonclassicality that we call as the process output nonclassicality for multimode quantum channels. The motivation comes from an information-theoretic point of view and the emphasis is on the output states of a channel.  We deem a channel to be `classical' if its outputs are always classical irrespective of the input, i.e., if the channel is nonclassicality breaking, and nonclassical otherwise. Our condition is stronger than the one considered by Rahimi-Keshari et al. [PRL {\bf 110}, 160401 (2013)] and we compare the two approaches. Using our framework we then quantify the nonclassicality of a quantum process by introducing a noise-robustness type of measure that we call as the nonclassicality depth of a channel. It characterizes a certain threshold noise beyond which a given channel outputs only classical states. We achieve this by generalizing a prescription by Lee [PRA {\bf 44}, R2775 (1991)] to multimode states and then by extension to multimode channels.  
\end{abstract}
\maketitle

\section{Introduction}
Nonclassical properties of radiation fields have been of great interest both from a theoretical and experimental point of view\,\cite{knight-rev,dodo-rev}.  Nonclassical states play an important role in quantum-optical settings since they are a useful resource for implementing quantum information and communication protocols, examples include entanglement generation\,\cite{knight02,wang02,wolf03,nha09,ivan11,ivan12,caves13,vogel14,vogel14b}, and superactivation of quantum capacity of quantum channels\,\cite{smith11,superact}. Much effort has also been devoted to various experimental aspects of nonclassical states \,\cite{polzik08,wolf08,murr11,brooks12,bowen13,regal13,grote13,leuchs-rev}. Practically relevant examples of nonclassical states include squeezed coherent states, Fock states, anti-bunched states, noon states, cat states, to mention a few. 
 
A significant effort has been directed to the detection of nonclassicality of states \cite{dodo-rev}. 
Apart from the qualitative assessment of nonclassicality of states,  various measures of nonclassicality have also been proposed and studied in literature\,\cite{vogel14b,hillery87,lee91,lut95,twamley96,dodonov00,baseia03,reno03,marian04,zyc04,asboth05,vogel12,hamar13,nori15,nori15b}. 

Analogous to the characterization of nonclassicality of states it is useful to assess the qualitative and quantitative  nonclassical properties of quantum processes. One such approach was presented by Rahimi-Keshari et al. \cite{keshari13} where the authors define a classical channel as one that maps input coherent states to classical states and nonclassical if at least one input coherent state is mapped to a nonclassical state. 

In this article we take a different viewpoint where the emphasis is on the output states of a channel. The physical motivation behind this approach is the usefulness of a quantum channel in a quantum information protocol. If the protocol depends on resources of the nonclassical type, then a channel which outputs only classical states could be rendered redundant. The same is also true if one is interested in nonclassicality benchmarks for quantum sources. Hence, a systematic approach to demarcate those channels that output only classical states and those that do not serves to be useful. 

Another motivation for the emphasis on the output states of a channel stems from the mathematical notion of distance between quantum channels which is computed using induced norms on the space of quantum channels. A natural way to construct a distance measure between two quantum processes is to compute a distance (defined on the set of density operators) between output states of the two channels with an optimization over the input state space [or for example between the output states of the two channels tensored with the identity channel as in the case of the diamond norm] \cite{watrous}.

An important class of channels with the property that they output only classical states irrespective of the input are known as nonclassicality breaking channels. We wish to highlight that these channels are single-party notions since nonclassicality of states is also a single-party notion (and not like standard correlations that require at least  one other system for their definitions). For the useful class of bosonic Gaussian channels, the structure and properties of those that are nonclassicality breaking have been studied in \cite{kraus10,nb1,nbm}. 

We deem that a channel is classical if its outputs are always classical, i.e., the channel is nonclassicality breaking, and nonclassical otherwise. In other words, we base the notion of nonclassicality of a channel on what we call process output nonclassicality. It is within this framework that we define a measure of nonclassicality for quantum-optical channels. 

Noise effects are inevitable  in any realistic implementation of quantum information tasks and in general degrades nonclassicality\,\cite{gardiner,breuer}. Numerous manifestations of noise in quantum information protocols have previously been studied\,\cite{caves81,hall94,rossi03,lut07,cerf09,ralph11,robust,smith13,grosshans15}. It is the effect of noise on the nonclassical characteristics of a quantum system (both on states and output states of channels) that we exploit to define a measure of nonclassicality. 

The outline of the paper is as follows. In Section II we introduce our notion of classical and nonclassical channels and compare it with the approach of Rahimi-Keshari et al. In Section III we introduce a measure of nonclassicality for multimode states which we call as the nonclassicality depth. In Section IV we extend the framework of Sec. III to define a meaningful notion of nonclassicality depth for multimode channels. Finally, we conclude in Section V. 

\section{Process output nonclassicality}
In this section we introduce a demarcation of processes into classical and nonclassical ones based on the output characteristics of the process which we call as process output nonclassicality. 

We begin with a brief description of the notion of nonclassicality of states of continuous-variable systems. The division between classical and nonclassical states of continuous-variable systems manifests itself most transparently in quantum optics. The pioneering work of Glauber\,\cite{glauber} and Sudarshan\,\cite{ecg} laid the foundation for the phase-space description of classical and nonclassical states of light. 

Any state $\hr{}$ can be  described in terms of its Glauber-Sudarshan diagonal `weight' function $\phi(\bs{\alpha};\hr{})$\,\cite{ecg}\,:
\begin{align}
\hr{} = \int\frac{d^{2n} \bs{\alpha}}{\pi^n}\, {\phi} (\bs{\alpha};\hr{}) | \bs{\alpha} \rangle \langle \bs{\alpha} |.
\label{e1}
\end{align} 
Then $\hr{}$ is said to be classical if its associated $\phi(\bs{\alpha};\hr{})$ function is everywhere non-negative over the phase-space $\bs{\alpha} \in \mathbb{R}^{2n} \simeq \mathbb{C}^{n}$, else it is said to be nonclassical. In other words, a  classical state is  a convex mixture of coherent states, and coherent states are the most elementary of quantum states exhibiting classical behaviour\,\cite{wolfbook}. 

Analogously, there is a divide at the level of quantum processes between those that output only classical states and those that output at least one nonclassical state, in other words, between channels that are nonclassicality breaking and those that are not. Recall that a  channel $\Phi$ is said to be nonclassicality breaking if \,\cite{kraus10,nb1,nbm}
\begin{align}
\Phi[\hr{}] \text{ is classical } \forall ~ \hr{} \in \text{ state space}.
\end{align}
The structure and properties of single-mode and multimode nonclassicality breaking bosonic Gaussian channels  were studied in Refs.\,\cite{kraus10,nb1} and Ref.\,\cite{nbm} respectively. 
We wish to emphasize that nonclassicality breaking channels form a convex subset of the set of all channels just as classical states form a convex subset of the set of all states. 

{\em Process output nonclassicality}. As motivated in the Introduction, our approach to defining classical and nonclassical channels will lay emphasis on the output states of a channel. 

\begin{figure}
\centering
\scalebox{0.4}{\includegraphics{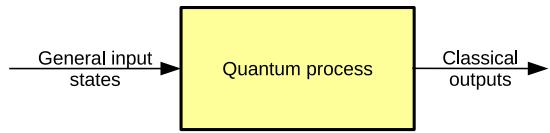}}
\caption{Showing the notion of nonclassicality breaking channels. A channel is said to be nonclassicality breaking if it outputs only classical states irrespective of the input state, i.e., it breaks the nonclassicality of input states. We deem that such channels are classical and nonclassical otherwise. \label{fig3}}
\end{figure}

We deem a channel to be classical if it outputs only classical states irrespective of the input, i.e. if the channel is nonclassicality breaking and this is schematically depicted in Fig. \ref{fig3}. Based on this approach, a nonclassical channel will be one for which there is at least one input state for which the corresponding output state is nonclassical. 

Further, for the case of bosonic Gaussian channels, we have a complete characterization of classical (nonclassicality breaking) channels \cite{nbm} that we will use later.

\subsection{Comparison with the work of Rahimi-Keshari et al.}
We now compare our approach to classical channels with that presented in the work by Rahimi-Keshari et al.\,\cite{keshari13}. There the authors introduced a notion of nonclassicality for a quantum process as the following: 

{\em Definition} \cite{keshari13}. A quantum process is nonclassical if it transforms an input coherent state to a nonclassical state. 

Further, in this framework, a classical process is one that maps input coherent states (and also classical states) to classical states. We note that their definition and subsequent discussion on the demarcation of processes into classical and nonclassical ones relies on the action of the channel on the set of coherent states. Examples of classical processes include all passive canonical transformations (that includes in particular the identity channel), photon subtraction, and the set of all nonclassicality breaking channels as to be explained below.

As mentioned in the previous subsection, we take an  approach in which the division of channels into classical and nonclassical ones depends on restrictions only on the output space for all inputs.  We call the method of Rahimi-Keshari et al. as quantum process nonclassicality (QPN) and we call our approach as process output nonclassicality. We compare the two approaches in Table. \ref{table2}. We see that in both the cases the set of channels which are classical forms a convex subset of the set of all channels. 

\begin{figure}
\centering
\scalebox{0.5}{\includegraphics{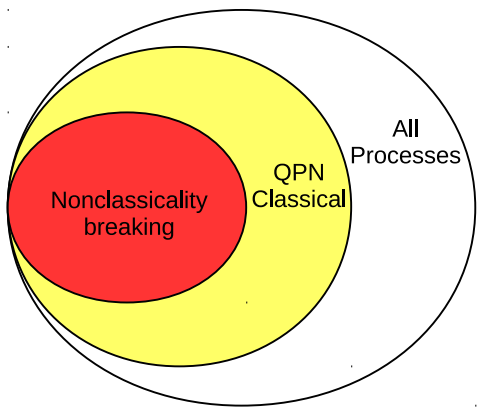}}
\caption{Showing the two notions of classical processes. The convex set labeled QPN-classical is the set of all processes that are classical according to the approach of Rahimi-Keshari et al. \cite{keshari13}. The set labeled nonclassicality breaking channels is the set of all classical channels as considered in this paper. We see that the set of nonclassicality breaking channels is a subset of the set of all QPN-classical channels (c.f. Table I). \label{fig2}}
\end{figure}

\begin{table*}
\begin{tabular}{clll}
&{\bf Property} &  {\bf Quantum process nonclassicality} & {\bf Process output nonclassicality}\\
\hline
\hline
1. &Classical process& Maps all input classical states to & Maps all input states to classical\\
&& classical states at the output& states at the output\\
& Examples & Photon subtraction, Identity channel, & All nonclassicality breaking channels \\
&& passive canonical transformations & like phase conjugation channels \\
2.& Convexity of the set of & QPN-classical processes form a convex ~  & Nonclassicality breaking channels form  a\\
&classical processes & subset of the set of all processes&  convex subset of the set of all processes\\
3.&Nonclassical process& Maps at least one input coherent state  & Maps at least one input state to a \\
&&to a nonclassical output state& nonclassical output state\\
&Examples &Cat-generation process & Quantum-limited amplifiers and \\
&&&  attenuators\\
4. &Quantum capacity for bosonic& Can be arbitrarily large  & Zero\\
&Gaussian channels that are `classical'~ &&\\
\hline
\end{tabular}
\caption{Showing a comparison of the two notions of classical processes. The first notion which we denote as quantum process nonclassicality (QPN) is the one introduced by Rahimi-Keshari et al. \cite{keshari13}. The second notion is the one introduced in this article and we call this notion as process output nonclassicality (PON) to highlight the fact that it is the outputs of the channels that are of importance in this approach. 
We note that in both the schemes the corresponding sets of classical channels form a convex subset of the set of all processes.  Further, we see that the set of classical channels according to the PON scheme is obtained by imposing a stronger condition than the one for the QPN scheme and is therefore a convex subset of the latter as depicted in Fig.\,\ref{fig2}.}
\label{table2}
\end{table*}

However, one important difference arises when we consider bosonic Gaussian channels, a ubiquitous and natural class of channel for continuous-variable systems, that are classical. We see that in the QPN approach, the quantum capacity can be arbitrarily large as can be seen for passive canonical transformations which also includes the identity channel. 

But in our approach we see that for nonclassicality-breaking (classical) bosonic Gaussian channels, the quantum capacity is strictly  zero  and the classical capacity is additive \cite{nbm}. 

The second comment is that our framework imposes a stronger condition on the description of classical channels as compared to the QPN approach. Hence  the set of nonclassicality breaking (i.e., classical channels) in our approach is a subset of the set of all classical channels in the QPN framework as depicted in Fig.\,\ref{fig2}. 

In the succeeding sections we shall use our framework to quantify the nonclassicality of quantum channels by defining a measure  that we call as the nonclassicality depth. As a foundation, we first develop the notion of nonclassicality depth for multimode states using a general noise-robustness framework in the next Section.

\section{nonclassicality depth of multimode states}
In this section we introduce a noise-robustness type of measure which we call as the nonclassicality depth of a multimode state. The approach is primarily motivated by the work of Lee \cite{lee91}.  Our framework is a generalized version not only in the single-mode case but also allows for suitable and straightforward generalization to the multimode case. Our main question is the following: what is the amount of noise that when added to a state renders it classical? We make this statement precise in the following paragraphs. 

\subsection{Noise model and minimal noise matrices}
We are interested in the effect of the additive classical noise channel, denoted by ${\cal B}_2[Y]$,   on the nonclassicality of states. The channel ${\cal B}_2[Y]$ is a bosonic Gaussian channel for which the corresponding $(X,Y)$ matrices are specified by the matrix pair $(1\!\!1, \, Y)$, where $Y$ is real, $Y=Y^T,\,Y \geq 0$, and the last inequality is also the complete positivity condition. 
The action of $\mathcal{B}_2(Y)$ on states can be described at the level of the symmetric-ordered characteristic function, denoted by $\chi_{W}(\bs{\xi};\hr{})$ with $\bs{\xi} \in \mathbb{R}^{2n}$, as\,\cite{werner01,caruso08}
\begin{align}
\chi_W(\bs{\xi};\hr{}^{\rm out}) = \chi_W(\bs{\xi};\hr{}^{\rm in})\,\exp\left[-\frac{1}{2}\bs{\xi}^T\,Y\,\bs{\xi} \right].
\label{e2}
\end{align}
We wish to point out that for the special case of additive classical noise channels, one can replace the symmetric-ordered characteristic function $\chi_{W}$ with any $s$-ordered characteristic function throughout in Eq.\,\eqref{e2}, leading to an equivalent description of the action of the noise channel ${\cal B}_2[Y]$.

Let us first denote the set of all noise matrices by $\mathbb{Y}$, i.e.,
\begin{align}
\mathbb{Y} := \left\{Y\,|\, Y \text{ real},\,Y=Y^T,\,Y\geq 0 \right\}.
\end{align}
So we have that $\mathbb{Y}$ is a partially ordered set with respect to the `$\geq$' operation. We  begin with the set of noisy channels whose action on a given state $\hr{}$ renders it classical. Let us denote this collection of corresponding noise matrices by $\mathbb{Y}[\hr{}]$, i.e.,
\begin{align}
\mathbb{Y}[\hr{}] = \left\{  Y \in \mathbb{Y} \,|\, \mathcal{B}_2[Y](\hr{}) \text{ is classical}\,\right\}.
\end{align}

We ask the following question\,: what is the minimal  additive classical noise channel to which if we input a state $\hr{}$  we get a classical state as output? This is useful because we would like to quantify the robustness of the nonclassicality of a given state $\hr{}$ against Gaussian noise. We wish to briefly introduce the notion of minimal noise matrix since it is an important concept we use repeatedly in this article and it appears in various scenarios. 

{\em Minimality}.   Let  ${\cal{C}}$ be an abstract condition that the elements of $\mathbb{Y}$ satisfy. We then say that $Y \in \mathbb{Y}$ satisfying ${\cal C}$ is `minimal' if it is one for which $Y - \Delta \in \mathbb{Y},\, \Delta \geq 0$ satisfying ${\cal C}$ implies $\Delta =0$.  It is clear that $\Delta$ can be taken to be a rank one projector, i.e., $Y$ is minimal in ${\cal C}$ if there does not exist a $|v \rangle$ such that $Y-\ket{v}\bra{v} \in \mathbb{Y}$ satisfies ${\cal C}$.   It is in this spirit that we  use the notion of minimal noise matrix and the associated $\mathcal{B}_2[Y]$ is deemed as the minimal noise channel. Depending on the condition $\mathcal{C}$ the minimal element may or may not be unique. 

As an example, let $\Gamma$ denote the set of all valid covariance matrices of an $n$-mode system. Let ${\cal C}$ be the condition that  the noise matrices are elements of $\Gamma$. Then the minimal elements of $\Gamma$ are given by matrices of the form $S^TS$ where $S \in {\rm Sp}(2n,\mathbb{R})$\,\cite{simon94}, the real symplectic group in $2n$ dimensions. \,$\square$

\subsection{Nonclassicality depth of states \label{ncdstates}}
Having introduced the noise model and the notion of minimal noise matrices, we now motivate our definition of nonclassicality depth of multimode states. We denote by $\mathbb{Y}_{\rm min}[\hr{}]$ the set of all minimal noise matrices for which the action of the corresponding classical noise channel on a state $\hr{}$ renders it classical, i.e.,
\begin{align}
\mathbb{Y}_{\rm min}[\hr{}] := \left\{ Y \,|\, Y \text{ is minimal in } \mathbb{Y}[\hr{}]  \right\}. 
\end{align}
Further, we have that for any state $\hr{}$\,\cite{glauber} 
\begin{align} 
0 \leq Y  \leq  2 1\!\!1 ~\forall~ Y \in \mathbb{Y}_{\rm min}[\hr{}].
\label{e2b}
\end{align} 

Let us now pick one such minimal noise matrix $Y \in \mathbb{Y}_{\rm min}[\hr{}]$.
We would like to define the largest variance (or fluctuation) of a quadrature of $Y$ as the nonclassicality depth of $\hr{}$. But consider another state $\hr{}^{\,\prime}$ which is related to $\hr{}$ through conjugation by a passive canonical unitary $U[R]$, i.e., $\hr{}^{\,\prime} = U[R] \,\hr{}\,U[R]^{\dagger}$ where $U[R]$ is the unitary representation of the symplectic group on $n$ modes that induces a phase-space rotation  $R$ on the mode operators and 
\begin{align}
R \in K(n) :=  {\rm Sp}(2n,\mathbb{R}) \cap {\rm SO}(2n,\mathbb{R}),
\end{align} 
 where ${\rm SO}(2n,\mathbb{R})$ stands for the special orthogonal group in $2n$ dimensions. 

It is clear from Eq.\,\eqref{e2} that for $\hr{}^{\,\prime}$  the corresponding minimal noise matrix $Y^{\,\prime}$   is given by $Y^{\,\prime} =  R\,Y\,R^T$. As far as nonclassicality depth is concerned we demand that both $\hr{}$ and $\hr{}^{\,\prime}$ have identical nonclassicality depth, i.e., we would like to treat $Y$ and $Y^{\,\prime}$ on an equal footing, and by extension also for any $R \in K(n)$. Further, we impose that this must be true for each such minimal element $Y \in \mathbb{Y}_{\rm min}[\hr{}]$. So we would need to optimize over all minimal elements of $\mathbb{Y}_{\rm min}[\hr{}]$. 

Let us denote the nonclassicality depth of a state $\hr{}$ by $\zeta[\hr{}]$, and having motivated it in the preceding discussion, we define it  as 
\begin{align}
\zeta[\hr{}] := \min_{Y \in \mathbb{Y}_{\rm min}[\hr{}]}\, \max_{\left\{ R \in K(n),\,\mu\right\}} \left[R\,Y\,R^T \right]_{\mu \mu},
\label{e3}
\end{align}
where $[A]_{ij}$ denotes the $(ij)^{\rm th}$ element of a matrix $A$.
The above Eq.\,\eqref{e3} can equivalently be  rewritten as 
\begin{align}
\zeta[\hr{}] &= \min_{Y \in \mathbb{Y}_{\rm min}[\hr{}]} \, \max_{\left\{ R \in K(n),\ket{v} \right\}}\, \bra{v}\, R\,Y\,R^T\,\ket{v}.
\label{e4}
\end{align}
This expression can be simplified using the following theorem and we have

{\em Theorem 1}\,: The nonclassicality depth of a multimode state $\hr{}$ is the largest eigenvalue of a corresponding minimal noise matrix $Y $ optimized over the set $\mathbb{Y}_{\rm min}[\hr{}]$, i.e.,
\begin{align}
\zeta[\hr{}] = \min_{Y \in \mathbb{Y}_{\rm min}[\hr{}]} \, ||Y||_{\infty}, 
\label{e5}
\end{align}
where $||\cdot||_{\infty}$ is the spectral norm. \\
\noindent
{\em Proof}\,: As mentioned earlier, the `max' part of the  optimization in Eq.\,\eqref{e4} includes one over all symplectic rotations $R \in K(n)$. The group $K(n)$ enjoys a special property that its  action is transitive on $S^{2n-1}$\,\cite{simon94,simon95}, where the set of all unit vectors of  $\mathbb{R}^{2n}$  constitutes the unit sphere denoted by $S^{2n-1}$. Hence, the `max' part of the optimization in Eq.\,\eqref{e4} can be replaced by an optimization over the eigenvalues of $Y$.\,$\blacksquare$

We can easily infer from Eqs.\,\eqref{e2b} and\,\eqref{e5}  that the nonclassicality depth of any  state $\hr{}$ satisfies
\begin{align}
0\leq \zeta[\hr{}] \leq 2,
\label{e6}
\end{align}
taking the value $0$ for a classical state and $2$ for a maximally nonclassical state; coherent states being examples of the former and  photon number states being examples of the latter\,\cite{lut95}. Further, by construction, we see that the nonclassicality depth of a state is invariant under conjugation by passive canonical unitary transformations.

We have  introduced two quantities to characterize the nonclassicality of a given state $\hr{}$. The first is the set of associated minimal noise matrices $\mathbb{Y}_{\rm min}[\hr{}]$ that contains elements for which the corresponding Gaussian additive classical noise channel precisely renders the state classical, and the second is a nonclassicality measure $\zeta[\hr{}]$ extracted from the set $\mathbb{Y}_{\rm min}[\hr{}]$ which we define as the nonclassicality depth of the state $\hr{}$. 

To summarize, given a state $\hr{}$ we first obtain the set $\mathbb{Y}[\hr{}]$ of noise matrices corresponding to channels that render the state classical, then we compute its minimal elements $\mathbb{Y}_{\rm min}[\hr{}]$, and finally obtain $\zeta[\hr{}]$ after performing a suitable optimization as mentioned in Theorem 1.  

\subsection{Operational interpretation of nonclassicality depth \label{opde}}
We now present an operational interpretation for the nonclassicality depth $\zeta[\hr{}]$ that would turn out to greatly simplify its computation. Let us impose the condition that the noise matrices should treat all the quadratures of each of the modes on an equal footing. In other words we impose that $Y = \alpha 1\!\!1$. Let us denote by $\widetilde{\mathbb{Y}}$ the set of all such noise matrices that are proportional to identity, i.e.,
\begin{align}
\widetilde{\mathbb{Y}} := \left\{ Y \, | \, Y= \alpha 1\!\!1, \alpha \geq 0 \right\}. 
\end{align}
Note that  the partial order originally present in the set $\mathbb{Y}$ has now been lifted to a total order in $\widetilde{\mathbb{Y}}$. Let $\widetilde{\mathbb{Y}}[\hr{}]$ denote the set of noise matrices in $\widetilde{\mathbb{Y}} $ (or equivalently $\alpha$'s) such that the action of the corresponding classical noise channels on a given state $\hr{}$ renders the state classical, i.e.,
\begin{align}
\widetilde{\mathbb{Y}}[\hr{}] := \left \{  Y \in \widetilde{\mathbb{Y}}  \, | \,  \mathcal{B}_2[Y](\hr{}) \text{ is classical}\right\}.
\label{e6b}
\end{align} 
The minimal element for the set $\widetilde{\mathbb{Y}}[\hr{}]$ is now unique due to the total order and we denote the corresponding $\alpha$ by $\alpha_{\rm min}[\hr{}]$. By comparing Eqs.\,\eqref{e5} (Theorem 1) and\,\eqref{e6b}, we see that 
\begin{align} 
\alpha_{\rm min}[\hr{}] = \zeta[\hr{}].
\end{align}

Therefore, to compute the nonclassicality depth of a given state we can take the noise matrices to be proportional to identity and then find the minimal noise matrix. However, in such a case we lose the precise information of the actual minimal noise channels that render the state classical, and this could be of practical importance.  

\subsubsection{Example of multimode Gaussian states}
As a simple example, we evaluate the nonclassicality depth of multimode Gaussian states. The condition for a Gaussian state to be classical is given by\,\cite{simon94} 
\begin{align}
V_{\hr{}} \geq 1\!\!1,
\label{e6d}
\end{align}
where $V_{\hr{}}$ is the covariance matrix of a given state $\hr{}$. As seen in the preceding subsection, to evaluate the nonclassicality depth $\zeta[\hr{}]$ of $\hr{}$ we have to find the minimum $\alpha$ such that $V_{\hr{}} + \alpha 1\!\!1 \geq 1\!\!1$. So  we have that
\begin{align}
\zeta[\hr{}] &= {\rm max}~(0,1-x),\nonumber\\
x&= \text{smallest eigenvalue of } V_{\hr{}}.
\end{align}
We find that the nonclassicality depth of every multimode Gaussian state falls in the range $0 \leq \zeta[\hr{}] \leq 1$. 

In passing we make a simple observation. Let $V_{\psi}= SS^T$ denote the covariance matrix of a pure Gaussian state $|\psi\rangle$, where $S$ is an element of the symplectic group ${\rm Sp}(2n,\mathbb{R})$. Let $V_{\hr{\psi}}= SS^T + \delta,\,\delta\geq 0$ denote the covariance matrix of a mixed state $\hr{\psi}$ that is a noisy version of $|\psi\rangle$. Then it follows that 
\begin{align}
\zeta[\hr{\psi}] \leq \zeta[|\psi\rangle\langle \psi|].  
\label{e6e}
\end{align}
In other words, Eq. \eqref{e6e} is a reflection of the fact that noise degrades nonclassicality.

\subsection{Connection to Lee's nonclassicality depth} 
Lee's nonclassicality depth $\tau[\hr{}]$ for a single-mode state\,\cite{lee91} is defined as the minimum $\tau$ such that the Glauber-Sudarshan $\phi(\bs{\alpha};\hr{})$ quasiprobability corresponding to
$\chi^{\tau}_{W}(\bs{\xi};\hr{})$ which is given by   
\begin{align}
\chi^{\tau}_{W}(\bs{\xi};\hr{}) = \chi_{W}(\bs{\xi};\hr{})\,\exp \left[-\,\frac{\tau}{2} |\bs{\xi}|^2 \right],
\label{e7}
\end{align}
is everywhere non-negative over the phase-space $\bs{\alpha} \in \mathbb{R}^{2} \simeq \mathbb{C}$, and it is also known that $0\leq \tau \leq 2$. We wish to highlight that the original formulation was stated in terms of quasiprobabilities (rather than at the level of the characteristic functions) and had a factor of $2$ incorporated in $\tau$ such that $0\leq \tau \leq 1$. But this particular way of expressing the nonclassicality depth of Lee provides for easy comparison with our general framework.

For the single-mode case, by comparing Eqs.\,\eqref{e2} and\,\eqref{e7}, we see that Lee's method is a special case where the noise matrix is restricted to being proportional to identity in our general framework. Then from the preceding discussion on the operational interpretation of nonclassicality depth we see that  our nonclassicality depth $\zeta$ coincides with $\tau$. By Theorem 1, we see that a similar situation arises if we simply generalize Eq.\,\eqref{e7}, i.e., Lee's prescription, to the multimode case. 

We see that our formalism  generalized Lee's prescription of the noise model for the single mode case and also allowed for suitable generalization to the multimode case.  That a connection to Lee's nonclassicality depth  holds not only in the single-mode case but also in the multimode scenario was observed due to nontrivial properties of the group $K(n)$ of symplectic rotations as mentioned in Theorem 1.

\subsection{Nonclassicality depth of a set of states \label{ncdset}}
We now define a notion of nonclassicality depth of a collection of states.  Let $\mathcal{S} = \{\hr{1},\,\hr{2},\cdots \}$ denote a given set of states. We now follow the steps that were used to define the nonclassicality depth of a single state. We denote by $\mathbb{Y}[{\cal S}]$ the set of noise matrices whose action on each of the states of ${\cal S}$ renders it classical, i.e.,
\begin{align}
\mathbb{Y}[{\cal S}] := \left\{ Y \in  \mathbb{Y} ~|~ \mathcal{B}_2[Y](\hr{}) \text{ is classical} ~\forall~ \hr{} \in \mathcal{S}  \right\}.
\label{e8}
\end{align}
We note that by slight abuse of notion we have used the same symbol $\mathbb{Y}[\cdot]$ but with the argument suitably changed from $\hr{}$ to ${\cal S}$. Then, as in the earlier case, we denote the minimal elements of $\mathbb{Y}[{\cal S}]$ by $\mathbb{Y}_{\rm min}[{\cal S}]$, i.e.,
\begin{align}
\mathbb{Y}_{\rm min}[{\cal S}] = \left\{ Y \,|\, Y \text{ is minimal in } \mathbb{Y}[{\cal S}]\right\}.
\end{align}

Tracing the same steps that were used to define the nonclassicality depth of states, we immediately obtain the nonclassicality depth of the set of states $\mathcal{S}$ as 
\begin{align}
\zeta[{\cal S}] := \min_{Y \in \mathbb{Y}_{\rm min}[{\cal S}]} \, ||Y||_{\infty}.
\label{e9}
\end{align}
Further, we note from the operational interpretation of nonclassicality depth of a state as mentioned in Sec. \ref{opde} that one has for the nonclassicality depth of a set of states 
\begin{align}
\zeta[{\cal S}] = \max_{\hr{} \in {\cal S}} \,\zeta[\hr{}].
\label{e9b}
\end{align}
It turns out that the nonclassicality depth of a set of states is an intermediate step that would lead us to define a meaningful  notion of nonclassicality depth of a quantum channel.

\section{Nonclassicality depth of multimode channels}
We now have all the necessary tools to define a suitable notion of nonclassicality depth of a quantum-optical channel based on our framework of process output nonclassicality introduced in Sec. II and the notion of nonclassicality depth for states introduced in Sec. III. 

Let $\Phi$ be a quantum channel and let us denote the collection of output states of $\Phi$ by 
\begin{align}
\mathcal{S}[\Phi] := \left \{ \Phi[\hr{}] \,|\,\hr{} \in \text{state space}\, \right \}.
\end{align}
Then, as in the earlier subsection on nonclassicality depth of a set of states, we define 
\begin{align}
\mathbb{Y}[\Phi] := \left\{ Y \in  \mathbb{Y} ~|~ \mathcal{B}_2[Y]\,(\hr{}) \text{ is classical} ~\forall~ \hr{} \in \mathcal{S}[\Phi]  \right\}.
\end{align}

\begin{figure*}
\centering
\scalebox{0.4}{
\includegraphics{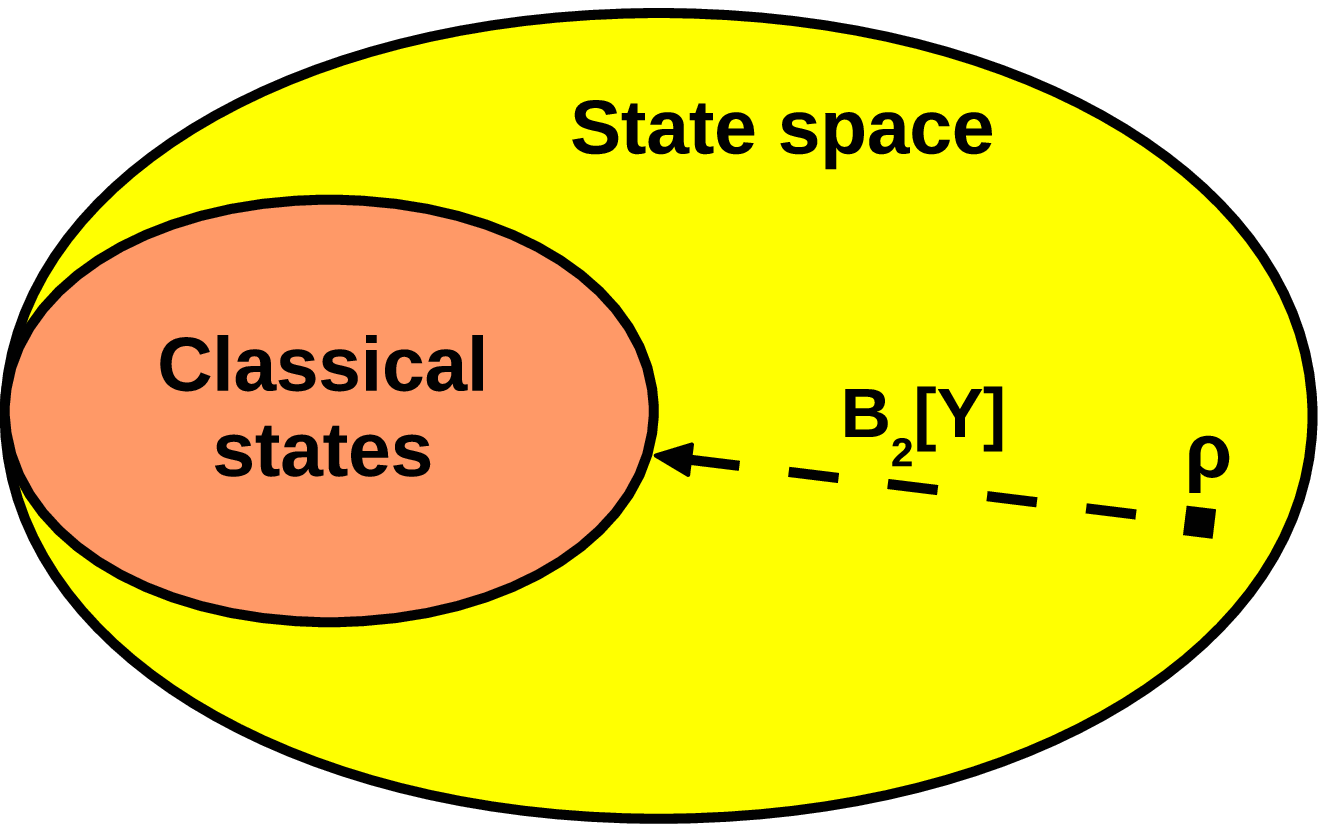}
\hspace{4cm}
\includegraphics{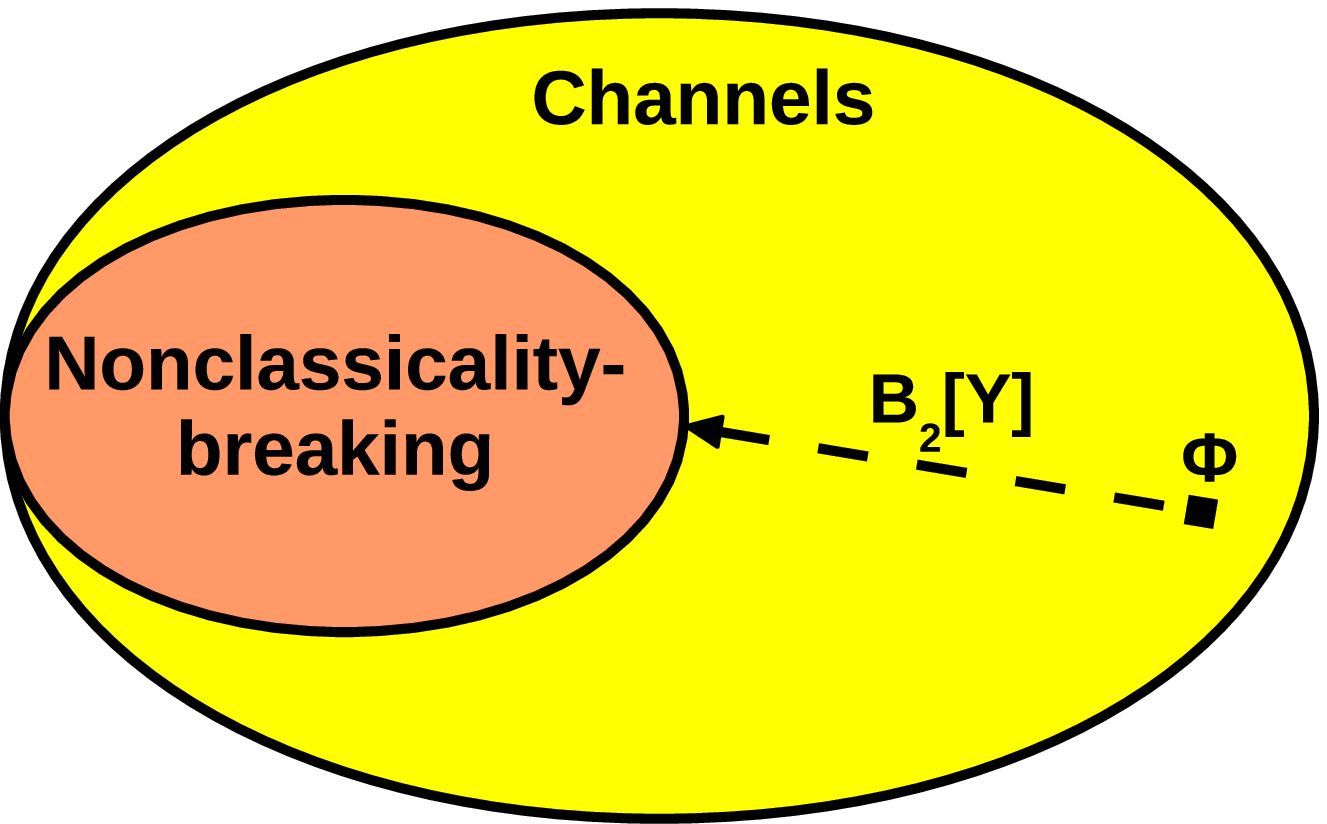}}
\caption{A schematic diagram depicting the respective notions of nonclassicality depth of a state and a channel. The figure on the left denotes the transition of a state $\hr{}$ from being  nonclassical to  classical  under the action of the additive classical noise channel $\mathcal{B}_2[Y]$. The figure on the right represents the transition of a general channel $\Phi$ that is not nonclassicality breaking (nonclassical) to a nonclassicality breaking (classical) channel under composition with a suitable $\mathcal{B}_2[Y]$. We wish to emphasize that the set of nonclassicality breaking channels is a convex set inside the set of all channels analogous to the convex set of classical states inside the entire state space. The corresponding nonclassicality depths of $\hr{}$ and $\Phi$ are denoted by $\zeta[\hr{}]$ and $\zeta[\Phi]$ respectively.  \label{fig}}
\end{figure*}

In other words, $\mathbb{Y}[\Phi]$ is the set of all noise matrices such that the corresponding noise channel action renders the channel $\Phi$  nonclassicality breaking. Recall from Eq.\,\eqref{e2} that a nonclassicality breaking (NB) channel is one for which the output state corresponding to every input state is classical. So equivalently we have  that
\begin{align}
\mathbb{Y}[\Phi] = \left\{ Y \in  \mathbb{Y} ~|~ \mathcal{B}_{2}[Y] \circ \Phi  \text{ is NB}\right\}.
\label{e10}
\end{align}
As in the previous sections we denote by $\mathbb{Y}_{\rm min}[\Phi]$ the minimal elements of $\mathbb{Y}[\Phi]$ and the nonclassicality depth of a quantum channel $\Phi$ by $\zeta[\Phi]$, and we have that
\begin{align}
\zeta[\Phi] := \min_{Y \in \mathbb{Y}_{\rm min}[\Phi]} ||Y||_{\infty}. 
\label{e11}
\end{align}
To briefly summarize, we provide a schematic representation of the procedure to compute the nonclassicality depth of states and channels in Fig.\,\ref{fig}. We note that our measure $\zeta[\hr{}]$ ($\zeta[\Phi]$) is faithful in the sense that it is zero for classical states (nonclassicality breaking channels) and nonzero otherwise. The measure is also bounded as can be seen from Eqs.\,\eqref{e2b} and\,\eqref{e6}. Finally, it also satisfies the appropriate symmetry properties as discussed in the next paragraph. 

We wish to note a subtlety in the invariance properties when transitioning  from nonclassicality depth of a state  (collection of states) to nonclassicality depth of a channel.  The nonclassicality depth of a state  (collection of states) is invariant under the action of an arbitrary passive canonical unitary transformation on the state (on all the states of the set). On the other hand, the nonclassicality depth of a channel is invariant under pre-processing the channel by an arbitrary unitary  and post-processing by an arbitrary passive canonical transformation. We restrict these pre-processing unitaries to canonical transformations when we are dealing with bosonic Gaussian channels (BGCs), our main example that we consider in the following subsections.

\subsection{Computing nonclassicality depth of single-mode BGCs}
As a simple illustration we compute the nonclassicality depth of single-mode quantum-limited bosonic Gaussian channels taken in its respective canonical forms and tabulate the same in Table\,\ref{table1}. For this purpose we make use of the explicit necessary and sufficient conditions for single-mode bosonic Gaussian channels to be nonclassicality breaking that were derived in\,\cite{nb1,nbm}. 

Since both the quantum-limited phase-conjugation channels and the singular channel are nonclassicality breaking\,\cite{kraus10,nb1} their respective nonclassicality depths evaluate to  $\zeta=0$. The same is also true for the  trivial case of $X=0$ which corresponds to the  $\kappa=0$ limit for the attenuator channel, where $\kappa$ is the corresponding attenuation parameter. For general attenuators we find that the nonclassicality depth  grows quadratically with  $\kappa$ and ultimately reaches the full strength of $2$ for $\kappa=1$, which corresponds to the identity channel. The last case of the quantum-limited amplifier channel requires the entire strength of the added noise with $\zeta$ evaluating to $2$ irrespective of the channel parameter.

\begin{table}
\begin{tabular}{lcc}
\hline
{\bf BGC} & $\bs{(X,Y)}$ & $\bs{\zeta [\Phi]}$  \\
\hline
\hline 
$C_1(\kappa)$\,: Attenuator  &$(\kappa\,\id{2}, (1-\kappa^2)\,\id{2})$; & $\zeta[\kappa]= 2\kappa^2$\\
&$0\leq \kappa \leq 1$&\\
$C_2(\kappa)$\,: Amplifier &$(\kappa\,\id{2}, (\kappa^2-1)\,\id{2})$; &$\zeta[\kappa]= 2$\\
&$ \kappa > 1$&\\
$D(\kappa)$\,: Phase-conjugation  &$(\kappa\,\sigma_3, (\kappa^2+1)\,\id{2})$; &$\zeta[\kappa]= 0$\\
&$ \kappa > 0$&\\
$A_2$\,: Singular &$((\id{2}+ \sigma_3)/2,\,\id{2})$&$\zeta[A_2]=0$ \\
\hline
\end{tabular}
\caption{Showing the nonclassicality depth of single-mode quantum-limited bosonic Gaussian channels (BGCs) taken in their respective canonical forms\,\cite{caruso06,holevo07}. The quantum-limited phase-conjugation and singular channels are all nonclassicality breaking\,\cite{kraus10,nb1}, and so have nonclassicality depth zero. The quantum-limited amplifier has the largest possible nonclassicality depth irrespective of the channel parameter  and the nonclassicality depth  of the quantum-limited attenuator is monotonically increasing with increasing channel parameter $\kappa$.}
\label{table1}
\end{table}
 
The main take-away from this computation of nonclassicality depth, as mentioned in Table\,\ref{table1}, is its reflection on the nonclassicality of states at the output of these channels. For the amplifier channel there exists at least one output which is maximally nonclassical irrespective of the channel parameter $\kappa$, for the attenuator channel the output with the maximum nonclassicality depth has nonclassicality depth which is monotonically increasing with the channel parameter $\kappa$ reaching the maximum value 2 for $\kappa=1$, and finally, both the phase-conjugation and the singular channel always only produce classical states, in other words, they are nonclassicality breaking. So we see how the nonclassicality depth of channels reflects in the nonclassical character of its output states. 

We wish to highlight that for the case of single-mode bosonic Gaussian channels we skipped the explicit optimization procedure and instead used the criterion for nonclassicality breaking\,\cite{nb1} for these channels which is known through other phase-space techniques. Also the nonclassicality depths for the noisy versions of these single-mode BGCs can also be evaluated in a similar way. We explain this directly for the multimode case that we consider next.

\subsection{Nonclassicality depth of multimode BGCs}
Let us consider an $n$-mode  bosonic Gaussian channel $\Phi(X,Y)$ described by the matrix pair $(X,Y)$, which satisfies the complete positivity condition\,\cite{verbure79,lindblad00,caruso08} 
\begin{align}
\mathcal{V}(X,Y) + i\Omega \geq 0,
\label{e12}
\end{align} 
where $\mathcal{V}(X,Y) = Y - iX^T\,\Omega\,X$ is the characteristic matrix associated with the matrix pair $(X,Y)$\,\cite{nbm}, and $\Omega = \oplus\, i \sigma_2 $, with $\sigma_2$ being the antisymmetric Pauli matrix, is the symplectic metric for an $n$-mode system. 

To obtain the nonclassicality depth, as discussed in Sec. \ref{opde}, we need to find the minimum $\alpha$ such that $(X,Y + \alpha 1\!\!1)$ is nonclassicality breaking. We recollect that for bosonic Gaussian channels the noise model of Eq. \eqref{e3} simply reflects as an additive noise at the level of the $Y$ matrix that represents the channel. 

So for computing the nonclassicality depth we resort to a criterion derived recently  in Ref.\,\cite{nbm} that tells us when a given bosonic Gaussian channel $\Phi(X,Y)$ is nonclassicality breaking\,:  
\begin{align}
\mathcal{V}(X,Y) &\geq 1\!\!1.
\label{e13}
\end{align}
So we need to find the minimal $\alpha$ such that 
\begin{align}
Y + \alpha 1\!\!1 - i X^T\,\Omega\,X \geq 1\!\!1.
\end{align}
By inspection we see that 
\begin{align}
\zeta[\Phi(X,Y)] &= {\rm max} \left(0, 1-x \right),\nonumber \\
x &= \text{ smallest eigenvalue of } \mathcal{V}(X,Y),
\label{e14} 
\end{align}
where $\zeta[\Phi(X,Y)]$ denotes the nonclassicality depth of the bosonic Gaussian channel corresponding to the matrix pair $(X,Y)$. 
We wish to note that unlike the case of Gaussian states, the nonclassicality depth of general multimode bosonic Gaussian channels can have any value in the range of $\zeta$ (Eq.\,\eqref{e6}), as demonstrated for the single-mode case in Table I. 

We can follow a similar procedure for bosonic Gaussian channels that are quantum-limited. Let $Y_{\rm QL}$ denote a minimal noise matrix satisfying Eq.\,\eqref{e12} for a given $X$. The bosonic Gaussian channel corresponding to the matrix pair $(X,Y_{\rm QL})$ is known as a quantum-limited channel. Then we can simply replace $Y$ by $Y_{\rm QL}$  in Eq.\,\eqref{e14} and then compute the corresponding nonclassicality depth. We make a simple observation that for a given $X$
\begin{align}
\zeta[\Phi(X,Y)] \leq \zeta[\Phi(X,Y_{\rm QL})],
\end{align}
the channel analogue of Eq.\,\eqref{e6e}.

However, if one is interested in the minimal noise channels whose action on a given channel renders it nonclassicality breaking, then one should evaluate the minimal elements of the set of noise matrices $Y^{\,\prime}$ which satisfy 
\begin{align}
\mathcal{V}(X, Y + Y^{\,\prime}) \geq 1\!\!1. 
\end{align} 

In passing we note that a byproduct of the  nonclassicality depth in Eq.\,\eqref{e14} is that it provides upper bounds for nonclassicality depths of multimode nonGaussian states which are possible outputs of the bosonic Gaussian channel corresponding to $(X,Y)$. \\

\section{Conclusions}
We have presented a framework for studying the qualitative and quantitative nonclassical characteristics of a quantum channel. Our approach relies on the notion of process output nonclassicality. We deemed channels that output only classical states (i.e., the set of nonclassicality breaking channels) as classical channels. Then any channel that outputs even a single nonclassical state is a nonclassical channel in our framework. 

We  compared our approach with the one introduced by Rahimi-Keshari et al. (QPN framework) where the authors proposed a notion of nonclassicality of channels based on their action on input coherent states. We then listed the various similarities and differences between the QPN framework and our framework in Table \ref{table2}. The condition imposed on the notion of classical channels in our framework is stronger than the one in the QPN framework. A main difference was noted for the case of quantum capacity of bosonic Gaussian channels that are classical in the two approaches, namely, that while it can be arbitrary in the QPN framework (as can be seen for the identity channel), in our case it is strictly zero.  

Once we established the qualitative part of assessing the nonclassical characteristics of a quantum channel,  we introduced a nonclassicality measure for channels. We achieved this by first constructing a unified approach to quantify the nonclassicality depth of states (channels) based on the robustness of nonclassicality of states (output states of the channel) against Gaussian noise for which it is rendered classical (nonclassicality breaking). The measure is faithful, bounded, and has the right symmetry properties. 
The method was inspired by Lee's prescription \cite{lee91} which we generalized for both the single-mode and multimode case, and by suitable extension to multimode channels. We then computed the nonclassicality depth of all multimode Gaussian states and channels. 

Our nonclassicality depth  $\zeta[\Phi]$ provides a quantitative characterization of the tolerance of a general quantum channel to the additive Gaussian noise before it is rendered nonclassicality breaking. This is relevant from an experimental point of view since nonclassicality is a useful resource for implementation of many quantum information protocols and noise is inevitable in any realistic setting. 

We believe that the tools and techniques presented here could have connections to various aspects of continuous-variable systems like quantum process tomography \cite{qpt1,qpt2} and also for constructing nonclassicality benchmarks for quantum sources. Finally, we wish to add that other measures of nonclassicality that have been mentioned in the Introduction can be used to define suitable measures of process nonclassicality following the methods outlined here.

\subsection*{Acknowledgments} The author acknowledges useful discussions with Andreas Winter and John Calsamiglia.  The author is supported by the ERC, Advanced Grant  ``IRQUAT'', Contract No. ERC-2010-AdG-267386, the Spanish MINECO  Project No. FIS2013-40627-P, and the Generalitat de Catalunya CIRIT, Project No. 2014-SGR-966.

\end{document}